\documentclass[12pt,technote,twoside]{IEEEtran}
\usepackage{latexsym,amssymb,epsfig}
\usepackage{supertabular}

\newcommand{\C}[0]{{\mathbf{C}}}

\newcommand{\Z}{{\mathbf{Z}}}
\newcommand{\mat}[4]{\left(\begin{array}{rr}#1 & #2\\ #3 & #4 \end{array}\right)}

\newcommand{\nix}[1]{}

\def\qed{\quad{$\Box$}}

\def\trace{\mathop{{\rm tr}}\nolimits}
\def\ker{\mathop{{\rm ker}}\nolimits}

\newtheorem{theorem}{Theorem}

\begin{document}
\title{Beyond Stabilizer Codes I: Nice Error Bases}
\author{%
Andreas Klappenecker,
\thanks{
A.K.~thanks the Santa Fe Institute for support through their
Fellow-at-Large program, and the European Community for support
through the grant IST-1999-10596 (Q-ACTA).  M.R.~thanks the DFG for
support through the Graduiertenkolleg GRK-209/3-98.}
\thanks{
A.~Klappenecker is with the Department of Computer Science, 
Texas A\&M University, College Station, TX 77843-3112, USA (e-mail: klappi@cs.tamu.edu).}%
Martin R\"otteler 
\thanks{M.~R\"otteler is with the Institut f\"ur Algorithmen und
Kognitive Systeme, Forschungsgruppe Quantum Computing
(Professor Thomas Beth), Universit\"at Karlsruhe, Am Fasanengarten 5,
D-76\,128 Karlsruhe, Germany (e-mail: roettele@ira.uka.de).}%  
}
%\date{23. October 2000}
\maketitle

\renewcommand{\baselinestretch}{1.66}
\small\normalsize

\begin{abstract}
\noindent Nice error bases have been introduced by Knill as a
generalization of the Pauli basis. These bases are shown to be
projective representations of finite groups.  We classify all nice
error bases of small degree, and all nice error bases with abelian
index groups. We show that in general an index group of a nice error
basis is necessarily solvable.
\end{abstract}

\section{Introduction}
Errors resulting from imperfect gate operations and decoherence are
serious obstructions to quantum computing. Recent progress in quantum
error control and in fault tolerant computing gives hope that large
scale quantum computers can be
built~\cite{calderbank98,gottesman96,shor95,shor96,steane96}. Although
arbitrary error operators $E$ might affect the state of a qubit, it is
always possible to keep track of the error amplitudes by expressing
$E$ in terms of an error operator basis, such as
\begin{equation}\label{pauli}
\begin{array}{r@{\quad}r}
\rho(0,0)=\mat{1}{0}{0}{\phantom{-}1},&
\rho(0,1)=\mat{0}{\phantom{-}1}{1}{0},\\[2.5ex]
\rho(1,0)=\mat{1}{0}{0}{-1},&
\rho(1,1)=\mat{0}{-1}{1}{0}.
\end{array}
\end{equation}
This is an example of an error basis for a two-dimensional quantum
system. The purpose of this note is to study error bases for higher
dimensional quantum systems. There are many possible bases for the
algebra of $n\times n$-matrices. A particularly useful class of
unitary error bases -- called nice error bases -- has been introduced
by Knill in~\cite{knill96a,knill96b}.  The nice error bases are the
pillar of quantum error control
codes~\cite{knill96b,ashikhmin00,rains99} which generalize the
stabilizer codes~\cite{calderbank98,gottesman96}.

Note that the error operators in (\ref{pauli}) are parametrized by
elements of the Klein four-group $\Z_2\times \Z_2$.  In general, nice
error bases are also parametrized by a group. The group structure
underlying the bases turns out to be important in quantum error
control applications and in fault-tolerant quantum computing.

The purpose of this note is to give detailed information about nice
error bases of small degree. To save space, we do not list the bases
themselves. Instead, we focus on the group structure underlying the
bases. We will show that nice error bases are projective
representations of finite groups. Thus, the know\-ledge of the groups
is sufficient to construct the corresponding error bases.  We provide
the programs for this construction at the web site\\
\centerline{\texttt{http://www.cs.tamu.edu/faculty/klappi/ueb/ueb.html}}
This site contains a complete list of nice error bases up to
degree~11.

\section{Error Bases}
Let ${G}$ be a group of order $n^2$ with identity element~1.  A
\textit{nice error basis}\/ on ${\cal H}=\C^n$ is a set ${\cal E}=\{
\rho(g)\in {\cal U}(n) \,|\, g\in {G}\}$ of unitary matrices such that
\begin{tabbing}
i)\= (iiiii) \= \kill
\>(i)   \> $\rho(1)$ is the identity matrix,\\[1ex]
\>(ii)  \> $\trace\rho(g)=n\,\delta_{g,1}$ for all $g\in {G}$,\\[1ex]
\>(iii) \> $\rho(g)\rho(h)=\omega(g,h)\,\rho(gh)$ for all $g,h\in
{G}$,
\end{tabbing}
where $\omega(g,h)$ is a nonzero complex number depending on $(g,h)\in
G\times G$;  the function
$\omega\colon G\times G\rightarrow \C^\times$ is called the factor
system of $\rho$.
We call $G$ the \textit{index group}\/ of the error basis
${\cal E}$.

We discuss some consequences of the definitions.  The conditions (i)
and~(iii) state precisely that $\rho$ is a projective representation
of the index group~${G}$, see~\cite[p.~349]{curtis62}.  Projective
representations have been introduced by Issai Schur at the beginning
of the last century. In this note, we will make profitable use of some tools
developed by Schur and others. We refer the reader to~\cite{curtis62} for
a gentle introduction to projective 
representations of finite groups.

Condition (ii) ensures that the matrices $\rho(g)$ are pairwise
orthogonal with respect to the trace inner product $\langle A,B
\rangle = \trace(A^\dagger B)/n$, hence they form a basis of the
matrix algebra $M_{n}(\C)$. It follows that $\rho$ is an irreducible
projective representation. In other words, $\{0\}$ and ${\cal H}$ are the only
subspaces of the vector space ${\cal H}$ that remain invariant under
the action of the representing matrices $\rho(g)$.

Furthermore, condition (ii) implies that the projective representation
$\rho$ is faithful, which simply means that $\rho(g)$ is not a scalar
multiple of the identity matrix unless $g=1$.

\begin{theorem}\label{nice}
Let ${\cal E}= \{ \rho(g)\,|\,g\in {G}\}$ be a set of unitary matrices
parametrized by the elements of a finite group $G$.  The set ${\cal
E}$ is a nice error basis with index group ${G}$ if and only if $\rho$
is a unitary irreducible faithful projective representation of ${G}$
of degree $|{G}|^{1/2}.$
\end{theorem}
\proof If ${\cal E}$ is a nice error basis, then the above discussion
shows that $\rho$ is indeed a
unitary irreducible faithful projective representation of degree
$|{G}|^{1/2}$. Conversely, $\rho$ satisfies conditions (i)
and (iii), since it is a projective representation. The assumption on
the degree gives condition (ii) for $g=1$. The representation $\rho$
is supposed to be faithful, which means that 
$$ \ker \rho = \{ g\in G\,|\, \rho(g)= cI\;\mbox{for some}\;
c\in \C \} = \{
1\}.$$
Here, $I$ denotes the identity matrix. 
The extremal degree condition $\trace\rho(1)=|{G}|^{1/2}$
implies that $\trace\rho(g)=0$ holds for all $g\in G - \ker
\rho$, see Corollary 11.13 in~\cite[p.~79]{karpilovsky94}. It follows that the
representation $\rho$ satisfies condition (ii) for the nonidentity
elements as well.~\qed

As a consequence of this theorem, we obtain a complete classification
of all nice error bases with abelian index group. Recall that a group
$G$ is said to be of symmetric type if $G\cong H\times H$ for some
group~$H$.

\begin{theorem}\label{abelian}
If a nice error basis ${\cal E}$ has an abelian index group~${G}$,
then~${G}$ is of symmetric type.  Conversely, any finite abelian
group~${G}$ of symmetric type is index group of a nice error basis.
\end{theorem}
\proof Theorem~\ref{nice} shows that the error basis ${\cal E}$ can be
understood as a faithful, irreducible projective representation
of~$G$. It was already shown by Frucht \cite[XIII, p.~24]{frucht31} in
1931 that an abelian group admits a faithful irreducible projective
representation only if it is of symmetric type.

On the other hand, if $G$ is a finite abelian group of symmetric type,
then it has a faithful irreducible unitary projective
representation~$\rho$, again by result XIII
in~\cite[p.~24]{frucht31}. The degree of~$\rho$ is given by
$\sqrt{|G|}$, see result XII in~\cite[p.~22]{frucht31}.  Therefore,
the set of representing matrices $\{\rho(g)\,|\, g\in G\}$ is a nice
error basis according to Theorem~\ref{nice}.~\qed

\section{Abstract Error Groups}
The previous section characterized all nice error bases with abelian
index groups. The case of nonabelian index groups is more complicated.
We show that an index group of a nice error basis is a solvable group.
We derive this result by studying slightly larger groups known as 
abstract error groups.

Let $G$ be the index group of a nice error basis $\{\rho(g) \,|\, g\in
G\}$. We assume that the factor system $\omega$ is of finite order,
i.\,e., that there exists a natural number $m$ such that
$\omega(g,h)^m=1$ for all $g,h\in G$. This can always be achieved by
multiplying the representation matrices $\rho(g)$ with a suitable
phase factor if necessary~\cite{knill96b}. Denote by $T$ the cyclic
group generated by the values of $\omega$.  Define an operation
$\circ$ on the set $H=T\times G$ by
$$ (a,g)\circ(b,h) = (ab\,\omega(g,h), gh), \qquad a,b\in T, g,h\in
G.$$ It turns out that $H$ is a finite group with respect to this
multiplication, the $\omega$-covering group of $G$
\cite[p.~134]{karpilovsky93}. A group isomorphic to such an
$\omega$-covering group of an index group of a nice error basis was
called \textit{abstract error group}\/ by Knill.

Given a nice error basis $\{\rho(g) \,|\, g\in G\}$, then the abstract
error group is isomorphic to the group generated by the matrices
$\rho(g)$. The assumption that the factor system $\omega$ is of finite
order ensures that the abstract error group is finite. For instance,
if we take the $n$-fold tensor product of matrices~(\ref{pauli}), then
we obtain a nice error basis for a system of $n$ qubits.  The abstract
error group $H$ generated by these matrices is isomorphic to a
so-called extraspecial $2$-group, cf.~\cite{calderbank98}.

\begin{theorem}
A group $H$ is an abstract error group if and only if $H$ is a group
of central type with cyclic center $Z(H)$. In particular, all abstract
error groups are solvable groups.
\end{theorem}
 
\proof Recall that a group $H$ is said to be of central type if and
only if there exists an irreducible ordinary character $\chi$ of $H$
with $\chi(1)^2=(H:Z(H))$.

\begin{enumerate}
\item[\textit{Step 1.}] 
If $H$ is an abstract error group, then it is (isomorphic to) an
$\omega$-covering group of an index group $G$, which has an
irreducible projective representation of degree $|G|^{1/2}$ with
factor system~$\omega$. In particular, $G\cong H/T$ for some cyclic
central subgroup $T$ of~$H$.  Each irreducible projective
repesentation of $G$ with factor system~$\omega$ lifts to an ordinary
irreducible representation of $H$ of the same degree.

Consequently, $H$ has an irreducible ordinary character $\chi$ with
$\chi(1)^2=(H:T)$. Each irreducible character $\chi$ of a group $H$
satisfies the inequality $\chi(1)^2\le (H:Z(H))$, whence $T=Z(H)$. It
follows that $H$ is a group of central type with cyclic center.

\item[\textit{Step 2.}]  
Conversely, suppose that $H$ is a group of central type with cyclic
center.  It was shown in a seminal work by Pahlings~\cite{pahlings70}
that $H$ has a faithful irreducible unitary representation
$\varrho$ of degree $(H:Z(H))^{1/2}$; this can be obtained by combining
the Corollary and Proposition~5 in~\cite{pahlings70}.

Let $G=H/Z(H)$, and denote by $W=\{ x_g\,|\, g\in G\}$, with $1\in W$, a set of coset
representatives for $Z(H)$ in $H$. Define a projective representation
$\rho$ of $G$ by 
$$ \rho(g)=\varrho(x_g), \qquad g\in G.$$ This projective
representation $\rho$ is unitary, irreducible, and faithful.
Theorem~\ref{nice} shows that $G$ is an index group of a nice error
basis. Finally, $H$ is by construction isomorphic to an
$\omega$-covering group of $G$, hence an abstract error group.

\item[\textit{Step 3.}] 
A deep result by Howlett and Isaacs shows that all groups of central
type are solvable. The proof of this fact relies on the
classification of finite simple groups~\cite{howlett82}.~\qed
\end{enumerate}

The classification of all abstract error groups was posed as an open
problem by Knill~\cite{knill96b}. The previous theorem showed that all
abstract error groups are solvable. It is known that all solvable
groups can occur as {\em subgroups} of index groups of nice error
bases, see Theorem 1.2 in~\cite{gagola74}. On the other hand, it is
known that not all solvable groups can occur as index
groups~\cite{shahriari91}. This delicate situation makes it difficult
to find a complete characterizaton of index groups of nice error bases
and thus of abstract error groups. 

Unlike the general case, it is fairly simple to characterize the
abstract error groups with abelian index groups:
\begin{theorem}
A group $H$ is an abstract error group with abelian index group
$G\cong H/Z(H)$ if and only if $H$ is nilpotent of class at most 2
and $Z(H)$ is a cyclic group. 
\end{theorem}
\proof We denote by $H'$ the commutator subgroup
$$ H' = \langle [g,h]=g^{-1}h^{-1}gh \,|\, g,h\in H\rangle$$ 
of the group $H$. Recall that a quotient group $H/N$ of $H$ is abelian if and only if the normal subgroup $N$ of $H$ contains the commutator subgroup $H'$. 

Therefore, an abstract error group $H$ has an abelian index group $G\cong H/Z(H)$
if and only if $H'\subseteq Z(H)$. In other words, $H$ is nilpotent of
class at most 2. The center of $H$ is cyclic, since there exists a
faithful irreducible representation of $H$.

Conversely, if $H$ is nilpotent of class 2 with cyclic center, then it
has a faithful irreducible representation of degree
$\mbox{$(H\!:\!Z(H))^{1/2}$}$, cf.~the Corollary to Proposition~4
in~\cite{pahlings70}. This shows that $H$ is an abstract error
group. If $H$ is abelian, hence cyclic, then $H$ is an abstract error
group with trivial index group.~\qed

\section{Small Index Groups}
For practical purposes it is desirable to know the index groups of
nice error bases of small order. We determine all nonabelian index
groups of order 121 or less in this section.
Recall that all groups of order $p^2$, $p$ a prime, are
abelian. Therefore, the search for nonabelian index
groups of order $n^2$ can be restricted to composite numbers $n$.
We used two different methods for the computer search.
The first method searches the solvable groups of order $n^3$
to find abstract error groups, whence index groups appear as
factor groups. The second method determines the degrees of the
irreducible projective representations of potential index groups of
order $n^2$ by calculating the Schur representation group. 

\smallskip
\textit{Method 1.} 
For each index group $G$ of order $n^2$ there exists an abstract error
group $H$ of order at most $n^3$ such that $G\cong H/Z(H)$. Thus, we can
search the catalogue of solvable groups~\cite{gap} to find groups of central
type with cyclic center. A calculation of the character table
of all possible groups would be impractical. Here it is advantageous
to use the following criterion by Pahlings: {\em 
Let $H$ be a group with cyclic center $Z$, $G=H/Z$. Denote by
$\textup{Cl}_G(x)$ the conjugacy class of $x$ in $G$. 
The group $H$ is of central
type if and only if\/ $|\textup{Cl}_H(x)|>|\textup{Cl}_G(xZ)|$ 
for all $x\in H - Z$. }
This test allows to determine rapidly if a group is an abstract error
group or not. 

\smallskip\textit{Method 2.} 
In the second method we need to check whether a potential index
group $G$ of order $n^2$ has an irreducible projective representation
of degree~$n$. This can be done by calculating the Schur
representation group $G^*$ of $G$. This is a central extension of $G$
which has the particular feature that {\em all}\/ irreducible
projective representations of $G$ can be derived from the ordinary
representations of $G^*$. In particular, $G$ is an index group if and
only if there exists an irreducible ordinary representation of $G^*$ of degree
$n$.

\smallskip
The first method has the advantage that the test is rather quick, but
has the drawback that many groups have to be tested.  Using this
method we were able to determine all nonabelian index groups $G$ of
order $n^2$ with the exception of the case $n=8$.  There exist 10 494
213 groups of order $n^3=8^3=512$ so that it is not feasible to search
for abstract error groups of this size. In contrast, the second method
has to search only 256 potential nonabelian index groups of order
$n^2=64$.  The calculation of the Schur representation group and its
ordinary representations of each candidate showed that there are $39$
nonabelian index groups of order $8^2$. The number of nonabelian index
groups of order $n^2$ are summarized in Table~\ref{tab:nig} for small
values of $n$. The corresponding index groups are tabulated in the Appendix.
\begin{table}[h]
\caption{This table lists the number of Nonabelian Index Groups
(NAIGs) of nice error bases of degree $n$ for all $n<12$.} \label{tab:nig}
\begin{center}
$\begin{tabular}{|c|c|c|c|c|c|c|c|c|c|c|c|}
\hline
\mbox{Degree} n & 1 & 2 & 3 & 4& 5& 6& 7& 8& 9& 10 & 11 \\
\hline 
\# \mbox{NAIGs} & 0 & 0 & 0 & 2 & 0 & 2 & 0 & 39 & 3 & 1 &
0\\
\hline
\end{tabular}$
\end{center}
\end{table}

\section{Example}
We construct in this section a family of abstract error groups
$H_n$. The example illustrates how to decide whether a group is an
abstract error group or not. Let $H_n=\langle \tau, \alpha\rangle$ be the group generated by 
composition of the maps 
$$ \tau = (x\mapsto x+1\bmod 2^n), \qquad \alpha = (x\mapsto 5x\bmod
2^n).$$ Notice that $H_n=A\rtimes B$ is a semidirect product of the
groups $A=\langle \tau\rangle$ and $B=\langle \alpha\rangle$, that is,
$A$ is a normal subgroup of $H_n$ and the factor group $H_n/A$ is
isomorphic to $B$.

\begin{theorem}
The group $H_n$ is an abstract error group of order $2^{2n-2}$. 
The index group $H_n/Z(H_n)$ is nonabelian provided that $n\ge 5$. 
\end{theorem}
\proof We show in the first part of the proof that the center
$Z=Z(H_n)$ of $H_n$ is cyclic of order 4. Then we show that $H_n$ is a
group of central type. Here we use the criterion by Pahlings in the
following form: the conjugacy class of any noncentral element
contains two distinct elements, which are in the same coset modulo the
center $Z(H_n)$.

\begin{enumerate}
\item The elements of the group $H_n$ can be written in the form
$$ H_n=\{ \tau^k\alpha^l\, |\, 0\le k< 2^n, 0\le l< 2^{n-2}\},$$ since
the relation $\alpha\tau=\tau^5\alpha$ holds.  
\item An element
$\tau^k\alpha^l$ of $H_n$ is in the center if and only if it coincides
with all its conjugates. Comparing 
$$ \tau^{-1}\tau^k\alpha^l\tau=(x\mapsto 5^lx+5^l+k-1\bmod 2^n)$$ with
$\tau^k\alpha^l=(x\mapsto 5^lx+k\bmod 2^n)$ shows that $5^l\equiv
1\bmod 2^n$ needs to holds for elements in~$Z$. This implies
$l=0$, and thus all elements of $Z$ are in $A$.  Notice that
$$\alpha\tau^k\alpha^{-1}=(x\mapsto x+5k \bmod 2^n)$$ 
is equal to
$\tau^k=(x\mapsto x+k\bmod 2^n)$ if and only if
$4k\equiv 0\bmod 2^n$ holds.  Therefore, the center is given by 
$Z=\langle \tau^{2^{n-2}}\rangle$, a cyclic group of order~4.

\item Consider an element of the form 
$$a=\tau^k\alpha^l=(x\mapsto 5^lx+k\bmod 2^n)$$ 
with $1\le
l<2^{n-2}$. Conjugation with $\tau^m$ yields
$$ b=\tau^{-m}(\tau^k\alpha^l)\tau^m= (x\mapsto 5^lx+(5^l-1)m+k \bmod
2^n).$$
Choose the smallest $m$ with $0<m\le 2^{n-3}$ such that
$$(5^l-1)m\equiv 0 \bmod 2^{n-2}$$ 
holds. 
Then $a\equiv b\bmod Z$, but $a\neq b$.

\item Consider an element of the form 
$$a=\tau^k=(x\mapsto x+k\bmod 2^n).$$ 
Conjugation with $\alpha^m$ yields
$$ \alpha^{-m}\tau^k\alpha^m= (x\mapsto x+5^mk\bmod 2^n).$$
Recall that the subgroup generated by $5$ in the group of units
$(\Z/2^n\Z)^\times$ is of the form
$ \langle 5\rangle=\{ 1+4\Z\bmod 2^n\},$
see \cite[p.~72]{lang90}. Write $k$ in the form $k=k_2k_2'$, where $k_2$
is a power of 2 and $k_2'$ is not divisible by 2. 
Consequently, the conjugacy class $\textup{Cl}(a)$ of
$a$ is given by 
$$ \textup{Cl}(a)= 
\{ x+r \,|\, r\in k+4k_2\Z \bmod 2^n\},$$ 
since $x\mapsto k_2'x\bmod 2^n$ is a permutation of the set $4\Z/2^n\Z$.

If $|\textup{Cl}(a)|>1$, then the elements $\tau^k$ and $\tau^\ell$, with
$\ell=k+2^{n-1}\bmod 2^n$, are both in
$\textup{Cl}(a)$, and $\tau^k\equiv \tau^\ell\bmod Z$. 

The last two steps showed that in the conjugacy class of a noncentral
element there are always two distinct elements which are congruent
modulo $Z$.  Therefore, $H_n$ is of central type.
\item The factor group $H_n/Z$ is nonabelian for $n\ge 5$, since 
the maps 
$$\tau\alpha=(x\mapsto 5x+1\bmod 2^n)\quad\mbox{and}\quad
\alpha\tau=(x\mapsto 5x+5\bmod 2^n)$$ 
are not equivalent modulo $Z$, 
hence $\alpha Z\cdot \tau Z\neq \tau Z\cdot \alpha Z$.~\qed
\end{enumerate}
In some sense the groups $H_n$ are getting more and more complicated
for larger $n$. Indeed, it is not difficult to show that the group
$H_n$ is nilpotent of class $\lfloor (n+1)/2\rfloor$ for $n\ge 5$.  On
the other hand, the construction of an error basis for the index
group $H_n/Z(H_n)$ is rather simple.

For a semidirect product $A\rtimes B$ with abelian normal subgroup $A$
it is easy to write down the representations with the help of the
method of little groups by Mackey and Wigner~\cite[p.~146]{grove97}.
Let $\phi\colon \Z/2^n\Z\rightarrow \C$ be the function defined by
$\phi(x)=\exp(2\pi i\, 5^x/2^n)$, where $i^2=-1$.  Then the diagonal
matrix $\rho(\tau)=
\textup{diag}(\phi(0), \phi(1), \dots, \phi(2^{n-2}-1))$ and
the shift
$$ \rho(\alpha)=
\left(\begin{array}{cllc}
0 & 1 \\
  & \ddots&\ddots \\
&& \ddots & 1\\
1 &&& 0 
\end{array}\right)
$$
define a faithful irreducible unitary ordinary representation of $H_n$ of
degree $2^{n-2}$.  We obtain a nice error basis with nonabelian index group 
by 
$$ {\cal E} = \{\, 
\rho(\tau)^k\rho(\alpha)^\ell\,|\, 0\le k,\ell\le 2^{n-2}-1\,\}.$$

\section{Conclusions}
Nice error bases have found various applications in quantum
computing. The teleportation of quantum states \cite{bennett93} is
such an instance; the relation of teleportation schemes and unitary
error bases is discussed in~\cite{werner00}. Unitary error bases are
an essential tool in the construction of quantum error control codes.
The know\-ledge of the group structure is of particular importance here,
since Clifford theory provides a natural explanation of stabilizer
codes and their higher-dimensional
generalizations~\cite{calderbank98,gottesman96,steane96,knill96b}.
Nice error bases are also of interest in the theory of noiseless
subsystems~\cite{knill00,zanardi00}.  

\appendix
\section*{List of Nonabelian Index Groups}
\label{app} 
We list in this appendix all nonabelian index groups of nice error
bases which are of order 121 or less. The first column of
Table~\ref{tab:list} denotes the degree $n$ of the faithful projective
representation of the index group~$G$. This index group~$G$ is then of
order~$n^2$. The second column gives the number of the group in the
Neub\"user catalogue used in MAGMA and GAP, cf.~\cite{magma,gap}. For
instance, $\texttt{G := SmallGroup(100,15)}$ gives the nonabelian
index group of degree $10$ in GAP. The isomorphism type of the group
is tabulated in the third column.

%\bibliographystyle{IEEEbib} 
%\bibliography{gates,ueb}

\begin{table}[hp]
\caption{List of all nonabelian index groups of nice error bases of degree less than $12$.}\label{tab:list}
\begin{center}
\epsfig{file=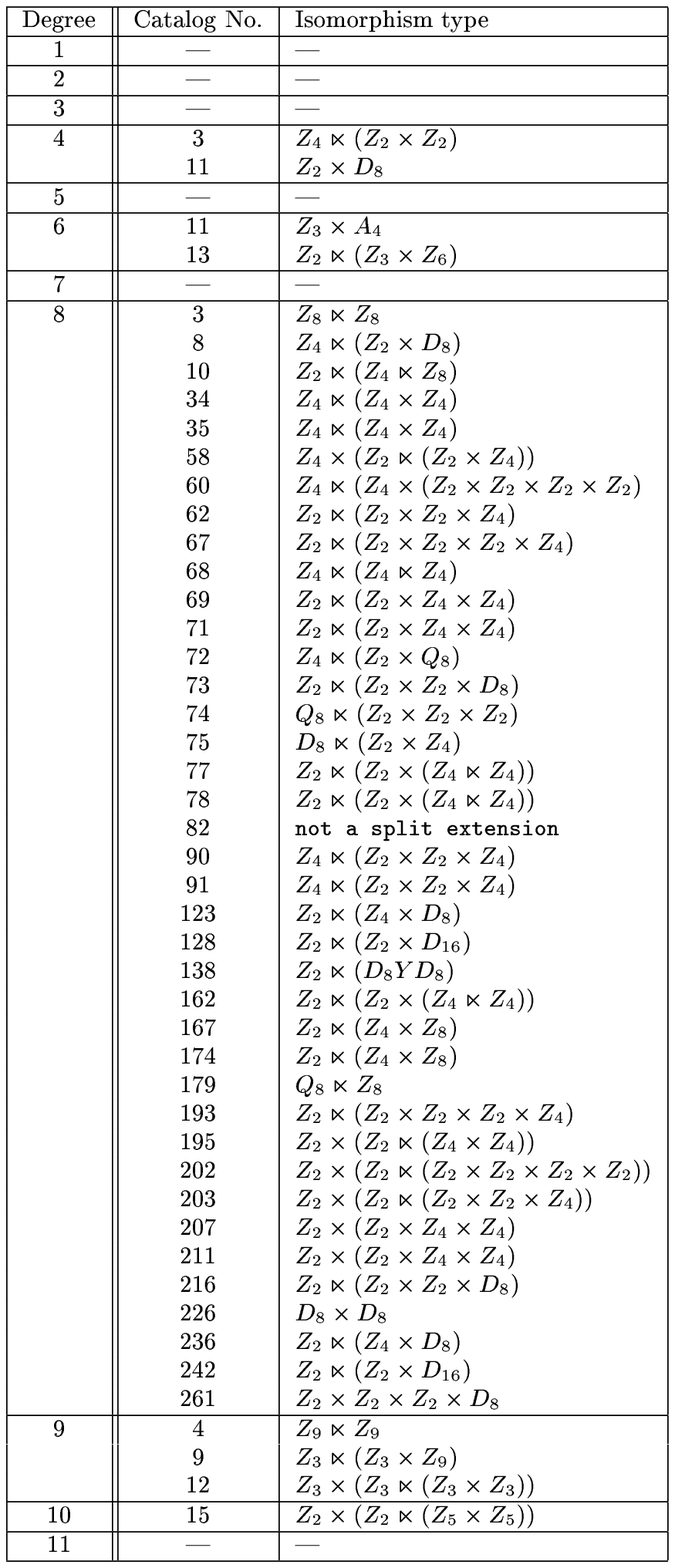}
\end{center}

\end{table}

\end{document}